\documentclass[journal]{IEEEtran}
\usepackage{times,amssymb,amsmath,epsfig}

\newcommand\nc{\newcommand}
\newtheorem{thm}{Theorem}
   \newtheorem{lem}{Lemma}
   \newtheorem{defi}{Definition}
   \newtheorem{remark}{Remark}
 \newtheorem{exa}{Example}

   \nc\bfa{{\bf a}}\nc\bfA{{\bf A}}\nc\cA{{\mathcal A}}
   \nc\bfb{{\bf b}}\nc\bfB{{\bf B}}\nc\cB{{\mathcal B}}
   \nc\bfc{{\bf c}}\nc\bfC{{\bf C}}\nc\cC{{\mathcal C}}
   \nc\bfd{{\bf d}}\nc\bfD{{\bf D}}\nc\cD{{\mathcal D}}
   \nc\bfe{{\bf e}}\nc\bfE{{\bf E}}\nc\cE{{\mathcal E}}
   \nc\bff{{\bf f}}\nc\bfF{{\bf F}}\nc\cF{{\mathcal F}}
   \nc\bfg{{\bf g}}\nc\bfG{{\bf G}}\nc\cG{{\mathcal G}}
   \nc\bfh{{\bf h}}\nc\bfH{{\bf H}}\nc\cH{{\mathcal H}}
   \nc\bfi{{\bf i}}\nc\bfI{{\bf I}}\nc\cI{{\mathcal I}}
   \nc\bfj{{\bf j}}\nc\bfJ{{\bf J}}\nc\cJ{{\mathcal J}}
   \nc\bfk{{\bf k}}\nc\bfK{{\bf K}}\nc\cK{{\mathcal K}}
   \nc\bfl{{\bf l}}\nc\bfL{{\bf L}}\nc\cL{{\mathcal L}}
   \nc\bfm{{\bf m}}\nc\bfM{{\bf M}}\nc\cM{{\mathcal M}}
   \nc\bfn{{\bf n}}\nc\bfN{{\bf N}}\nc\cN{{\mathcal N}}
   \nc\bfo{{\bf o}}\nc\bfO{{\bf O}}\nc\cO{{\mathcal O}}
   \nc\bfp{{\bf p}}\nc\bfP{{\bf P}}\nc\cP{{\mathcal P}}
   \nc\bfq{{\bf q}}\nc\bfQ{{\bf Q}}\nc\cQ{{\mathcal Q}}
   \nc\bfr{{\bf r}}\nc\bfR{{\bf R}}\nc\cR{{\mathcal R}}
   \nc\bfs{{\bf s}}\nc\bfS{{\bf S}}\nc\cS{{\mathcal S}}
   \nc\bft{{\bf t}}\nc\bfT{{\bf T}}\nc\cT{{\mathcal T}}
   \nc\bfu{{\bf u}}\nc\bfU{{\bf U}}\nc\cU{{\mathcal U}}
   \nc\bfv{{\bf v}}\nc\bfV{{\bf V}}\nc\cV{{\mathcal V}}
   \nc\bfw{{\bf w}}\nc\bfW{{\bf W}}\nc\cW{{\mathcal W}}
   \nc\bfx{{\bf x}}\nc\bfX{{\bf X}}\nc\cX{{\mathcal X}}
   \nc\bfy{{\bf y}}\nc\bfY{{\bf Y}}\nc\cY{{\mathcal Y}}
   \nc\bfz{{\bf z}}\nc\bfZ{{\bf Z}}\nc\cZ{{\mathcal Z}}
   \nc\bfzero{{\bf 0}}
   \nc\beq{\begin{equation}}
   \nc\eeq{\end{equation}}

   \nc\N{{\cal N}}

   \nc\barr{\begin{eqnarray*}}
   \nc\earr{\end{eqnarray*}}

\nc\ft{{\mathbb F}_2^{2t-1}}

\begin{document}

\title{ Boolean Functions, Projection Operators and Quantum Error Correcting
Codes}

\author{ Vaneet Aggarwal and Robert Calderbank  \\
{\small Department of Electrical Engineering, Princeton University,
NJ 08544, USA} \\  {\small Email: \{vaggarwa,
calderbk\}@princeton.edu}
\thanks{This work was supported in part by AFOSR under contract 00852833. The material in this paper was presented in part at the IEEE International Symposium on Information Theory, Nice, France, June 2007.} }%

\maketitle

\begin{abstract}
This paper describes a fundamental correspondence between Boolean functions and projection operators in Hilbert space. The correspondence is widely applicable, and it is used in this paper to provide a common mathematical framework for the design
of both additive and non-additive quantum error correcting codes. The new framework leads to the construction of a variety of codes including an infinite class of codes that extend the original $((5,6,2))$ code found by Rains \cite{Rains97}. It also extends to operator quantum error correcting codes.

\end{abstract}



\section{Introduction}

The additive or stabilizer construction of {\it quantum error
correcting codes} (QECC) takes a classical binary code that is
self-orthogonal with respect to  a certain symplectic inner
product, and produces a quantum code, with minimum distance
determined by the classical code (for more details  see
\cite{Calderbank98},  \cite{Calderbank97} and \cite{Gott97}). The first non-additive quantum error-correcting code was constructed by Rains {\it et al.} \cite{Rains97}. This code was
constructed numerically by building a projection operator with a
given weight distribution. Grassl and Beth \cite{Grassl97}
generalized this construction by introducing union quantum codes,
where the codes are formed by taking the sum of subspaces
generated by two quantum codes. Roychowdhury and Vatan
\cite{Roy98} gave some sufficient conditions for the existence of
nonadditive codes, and Arvind {\it et al.} \cite{Arvind02}
developed a theory of non-additive codes based on the Weyl
commutation relations. Most recently, Kribs {\it et al.}
\cite{Kribs05} introduced {\it operator quantum error correction} (OQEC)
which unifies the standard error
correction model, the method of decoherence-free subspaces, and
that of noiseless subsystems.

We will describe, what we believe to be the first mathematical framework for code design that
encompasses both additive and non-additive quantum error
correcting codes. It is based on a correspondence between
Boolean functions and projection operators in Hilbert space
that is described in Sections II and III. We have used an initial version of this correspondence to construct Grassmannian packings \cite{Ag06} and space-time codes for wireless communication \cite{Alexei2005}. However, the correspondence in Section III applies to a larger class of projection operators and includes the correspondence described in \cite{Alexei2005} as a special case (see Section IV). We note that prior work by Danielson \cite{thesis} interpreted Boolean functions as quantum states and developed a correspondence between Boolean functions and zero-dimensional quantum codes.

After introducing the fundamentals of quantum error correcting codes in Section V, we will derive in Section VI sufficient conditions for existence of QECC in terms of existence of certain Boolean function. This paper goes beyond deriving sufficient conditions, and constructs the quantum code if these properties are satisfied. Hence, we convert the problem of
finding a quantum code into a problem of finding Boolean
function satisfying certain properties. We also see how certain well-known codes fit into this scheme. We focus on non-degenerate codes
which is defensible given that we know of no parameters $k$, $M$ and $d$ for which there exists a (($k,M,d$)) degenerate QECC but
not a (($k,M,d$)) non-degenerate QECC (see \cite{aly}).
Further, in Section VII, we describe how this scheme fits into a
general framework of operator quantum error correcting codes. More precisely, we give sufficient conditions for the existence of $((k,M,N,d))$ stabilizer OQEC and also construct the code if these conditions are satisfied.

\section{Boolean Function}

A {\it Boolean function} is defined as a mapping $ f:\{ 0,1\} ^m \to \{
0,1\} $\cite{Preneel90}. The mapping $v = \sum\limits_{i = 1}^m {v_i 2^{i - 1}
} $ associates an integer $v$ from the set $\{ 0, 1, .... , 2^m-1 \}$ with a binary $m$-tuple $(v_m, ... , v_1 )$ with $v_i \in \{ 0,1 \}$. (Throughout the paper, $\sum$ represents addition over integers.) This integer is called the {\it decimal index} for a given
$m$-tuple.

An $m$-variable Boolean function $f$ can be specified by listing the
values at all decimal indices. The binary-valued vector of
function values $Y = [ y_0, y_1, ... , y_{2^m -1} ] $ is called
the {\it truth vector} for $f$.

An $m$-variable Boolean function $f(v_1 ,...,v_m )$ can be
represented as $ \sum\limits_{i = 0}^{2^m  - 1} {y_i v_1^{c_0(i) }
v_2^{c_1(i) } .... v_m^{c_{m - 1}(i) } } $ where $y_j$ is the
value of the Boolean function at the decimal index $j$ and
$c_0(j)$, $c_1(j)$, .... , $c_{m-1}(j) \in \{ 0, 1 \}$ are the
coordinates in the binary representation for $j$ (with $c_{m-1}$
as the most significant bit and $c_0$ as the least significant
bit) with $ v_j^{1} =v_j$ and $ v_j^{0} = \bar{v}_j$ (Theorem 7.7, \cite{lipschutz}).
\vspace{0.1in}

\begin{exa} The truth vector of the three-variable Boolean function $f(v_1,v_2,v_3) = v_1 v_2 \bar{v_3} $ is $Y =
[0 , 0
, 0 , 1 , 0 , 0 , 0 , 0 ]$
\end{exa}\vspace{0.1in}
\begin{defi}
The {\it Hamming weight} of a Boolean function is defined as the number of
nonzero elements in $Y$.
\end{defi}\vspace{0.1in}
\begin{defi} [\cite{Preneel90}]
Let $\oplus$ denote modulo two addition. The {\it (periodic) autocorrelation function} of a  Boolean function $f(v)$ at $a$ is
the inner product of $f$ with a shift of $f$ by $a$. More precisely,
$ r(a) = \sum\limits_{v = 0}^{2^m - 1} {( - 1)^{f(v) \oplus f(v
\oplus a)} } $ where  $a \in \{ 0 , 1, ... , 2^m - 1 \}$,   $ a =
\sum\limits_{i = 1}^m {a_i 2^{i - 1} } $. An autocorrelation
function is represented as a vector $R = [ r(0) , r(1) , ...
r(2^m-1) ]$
\end{defi}\vspace{0.1in}

\begin{defi}
The {\it complementary set} of a Boolean function $f(v)$ is defined by $ Cset_f  = \{
a|\sum\limits_{v = 0}^{2^m  - 1} {f(v)f(v \oplus a)}  = 0\} $
\end{defi}\vspace{0.1in}
This means that for any element $a$ in the $Cset_f$, $f(v) f(v
\oplus a)=0$ for any choice of $v \in  \{ 0,1,...,2^m - 1\}$.
The complementary set links distinguishability in the quantum world
(orthogonality of subspaces) with properties of Boolean
functions.  The quantity $f(v \oplus a)$ is the counterpart
in the quantum world of the quantum subspace after the error
has occurred, which is to be orthogonal to the original
subspace corresponding to $f(v)$ as will be described in later
sections. \vspace{0.1in}
\begin{lem}
If the Hamming weight of the Boolean function $f$ is $M$, and $M \le
2^{m-1}$ , then the complementary set $Cset_f  = \{ a|r(a) = 2^m  - 4M\} $

\end{lem}
\begin{proof}
 If $a$ $\in$ $Cset_f$ then $f(v)f(v \oplus a)
= 0 $ for all $v = 0, 1, ... , 2^m-1$ and the
supports of f(v) and $f(v \oplus a)$ are disjoint. Hence

\begin{eqnarray}
r(a) &=& \sum\limits_{v = 0}^{2^m - 1} {( - 1)^{f(v) \oplus f(v
\oplus a)} }\nonumber \\
&=&   (-1)^1 M + (-1)^1 M + (-1)^0 (2^m -2M)\nonumber \\
&=& 2^m - 4M \nonumber
\end{eqnarray}


Conversely suppose \[r(a) = \sum\limits_{v = 0}^{2^m - 1}
{( - 1)^{f(v) \oplus f(v \oplus a)} } = 2^m -4M.\] If the supports of $f(v)$, $f(v \oplus a)$ intersect in $N$ decimal indices then
\begin{eqnarray}
r(a) &=&  N - 2(M-N) + (2^m -2(M-N)-N)\nonumber \\
&=& 2^m - 4M +4N\nonumber
\end{eqnarray}

Hence, $N=0$ and $a
\in Cset_f$.  \end{proof}

\vspace{0.1in}
\begin{exa}
Let $f(v_1,v_2,v_3) = v_1 v_2 \bar{v_3} $. Then the vector B
corresponding to the autocorrelation function is
$[8,4,4,4,4,4,4,4]$, and $Cset_f = \{ 1 , 2 , 3 , 4 , 5 , 6 , 7 \}$.
\end{exa}

\section{Boolean Functions and a Logic of Projection Operators}
The authors of \cite{Alexei2005} connected Boolean logic to
projection operators derived from the Heisenberg-Weyl group. In
this section, we generalize these results to a
larger class of projection operators.

Let $\mathbb{B}(H)$ be the set of bounded linear operators on a
Hilbert space H. An operator $P \in \mathbb{B}(H)$ is called a
projection operator (sometimes we will use the terms orthogonal
projection operator and self-adjoint projection operator) on H iff
$P = PP^\dagger$. We denote the set of projection operators on H
by $\mathbb{P}(H)$ and the set of all subspaces of H by
$\mathbb{L}(H)$.
\vspace{.1in}
\begin{defi}
\begin{enumerate}
    \item If $
S \subseteq H$, the span of $S$ is defined as $
 \vee S =  \cap \{ K|K$ is a subspace in H with $S\subseteq K\}$.
 It is easy to see that $\vee S$ is the smallest subspace in H
 containing $S$.

    \item If $
S \subseteq H$, the orthogonal complement of $S$ is defined as  $ S^
\bot   = \{ x \in H| x \bot s\,$ for all $s \in S\}$.

    \item If $\mathbb{S}$ is a collection of subsets of H, we
    write $
 \vee _{S \in \mathbb{S}} S =  \vee ( \cup _{S \in \mathbb{S}} S)$.

\end{enumerate}
\end{defi}\vspace{0.1in}
\begin{defi}
Let $P \in \mathbb{P}(H)$ and let $ K = image(P) =\{Px|x\in
H\}$. We call $P$ the projection of $H$ onto $K$. Two
projections $P$ and $Q$ onto $K$ and $L$ are orthogonal
(denoted $P \bot Q$) if $PQ=0$. It is easy to verify that $ PQ
= 0 \Leftrightarrow K \bot L \Leftrightarrow QP = 0.$ (Theorem 5B.9, \cite{Cohen89})
\end{defi}\vspace{0.1in}
\begin{defi}
Let $P,Q$ $\in \mathbb{P}(H)$ with $K = image(P)$ and $L =
image(Q)$.
 Then \begin{itemize}
    \item $P<Q$ iff K$
 \subset$ L ($
K \ne L$ )

    \item $
P \vee Q$ is the projection of H onto $K \vee L$

    \item $
P \wedge Q $ is the projection of H onto $K \cap L$. \item
$\tilde P $ is the projection of H onto $K^\bot $.
\end{itemize}
\end{defi}\vspace{0.1in}
The structure $(\mathbb{P}(H), \leqslant , \bot )$ is a logic
with unit $I_H$ (identity map on $H$) and zero $Z_H$ (zero map on $H$) (Theorem 5B.18, \cite{Cohen89}). This
logic is called \textit{Projection Logic}.
\vspace{0.1in}
\begin{lem}[Theorem 5B.18, \cite{Cohen89}]
\label{pres} The map $ P \to image(P)$ from $\mathbb{P}(H)$ to
$\mathbb{L}(H)$ is a bijection that preserves order,
orthogonality, meet($\wedge$) and join($\vee$).

\end{lem}\vspace{0.1in}
\begin{lem}[\cite{Cohen89}]
\label{l1} If $<<P_k>>$ are pairwise orthogonal projection operators, in
$\mathbb{P}(H)$, then $ \vee _{k = 1}^\infty  P_k  = \sum\limits_{k
= 1}^\infty  {P_k }$.

\end{lem}\vspace{0.1in}
\begin{lem}[\cite{Cohen89}]\label{l2}
If  $P,Q \in \mathbb{P}(H)$, then\begin{enumerate}
    \item $PQ=QP$ iff PQ is a projection.
    \item If PQ is a projection, $image(PQ) = image(P) \cap
    image(Q)$.
\end{enumerate}

\end{lem}\vspace{0.1in}
\begin{lem}\label{l3}
If $P$ and $Q$ are commutative operators, then the distributive law
holds (and this law fails to hold for non-commutative operators).
Also, in this case,

\begin{enumerate}
  \item $P \wedge Q = PQ $
  \item $P \oplus Q \triangleq (P \wedge \tilde Q) \vee (\tilde P \wedge Q) = P + Q - 2PQ $
  \item $\tilde P = I - P$
  \item $P \vee Q = P + Q - PQ $
  \end{enumerate}

\end{lem}
\vspace{0.1in}
\begin{proof}
\begin{enumerate}
\item  From Lemma \ref{l2}, $image(PQ) = image(P) \cap
    image(Q)$. Hence, $image(PQ) = image(P \wedge Q)$ and by Lemma \ref{pres}, $P \wedge Q = PQ$.
\item We have \begin{eqnarray}
& P + Q - 2PQ  & = P(I - Q) + Q(I - P) \nonumber \\
& & \mathop =\limits^{(a)} [P(I - Q)] \vee [Q(I - P)] \nonumber \\
& & \mathop =\limits^{(b)} [P \wedge (I - Q)] \vee [Q \wedge (I - P)]\nonumber \\
& & \mathop =\limits^{(c)} P \oplus Q. \nonumber
\end{eqnarray}
where
$(a)$ follows from Lemma \ref{l1}, $(b)$ follows from Lemma \ref{l2} and $(c)$ follows directly from definition of $P \oplus Q$.
 \item $\tilde P = I - P$ follows
directly from Definition $6$.
\item We have \begin{eqnarray}
& (P \oplus Q) \vee (P \wedge Q)  & \mathop = \limits^{(d)} (P \oplus Q) + (P \wedge Q) \nonumber \\
& & \mathop = \limits^{(e)} P + Q - 2PQ + PQ \nonumber \\
& & = P + Q - PQ \nonumber\end{eqnarray}
Also, $(P \oplus Q) \vee (P \wedge Q)$
\begin{eqnarray} &   & =  (P \wedge \tilde Q) \vee (\tilde P \wedge Q)
\vee (P \wedge Q)  \nonumber \\ & &\mathop= \limits^{(f)} (P \wedge \tilde Q) \vee ((\tilde P \vee P) \wedge Q) \nonumber \\
& & = (P \wedge \tilde Q) \vee Q \ \  \nonumber \\
& & \mathop=\limits^{(g)} (P \vee Q) \wedge (\tilde Q \vee Q) \nonumber \\
& & =(P \vee Q) \nonumber
\end{eqnarray}
where $(d)$ follows from Lemma \ref{l1} since $P \oplus Q$ and $P \wedge Q$ are orthogonal ($(P+Q-2PQ)PQ = 0$), $(e)$ follows from Lemma \ref{l2}, and $(f)$, $(g)$ follows from the distributive laws. Hence, $P \vee Q = P + Q - PQ$.
\end{enumerate}
\end{proof}

\vspace{0.1in}
Next we define projection functions following
\cite{Alexei2005}.
\vspace{0.1in}

\begin{defi}\label{defi:bool}Given an arbitrary Boolean function $f(v_1, .... , v_m)$,
we define the {\it projection function} $f(P_1, ... , P_m)$ in which
$v_i$ in the Boolean function is replaced by $P_i$, multiplication
in the Boolean logic is replaced by the meet operation in the
projection logic,  summation in the Boolean logic (or the
\textit{or} function) is replaced by the join operation in the
projection logic and the not operation in Boolean logic is replaced by the
tilde ($\tilde P$) operation in the projection logic.
\end{defi}\vspace{0.1in}

As is standard when writing Boolean functions, we use \textit{xor} (modulo $2$ addition, represented by $\oplus$)
in place of \textit{or}, hence by above definition, we will replace
the \textit{xor} in the Boolean logic by the \textit{xor} operation
in the projection logic.

\vspace{0.1in}
\begin{thm}\label{thm:bool} If $(P_1, ..., P_m)$ are pairwise commutative
projection operators of dimension $2^{m-1}$ such that $P_1 P_2
.. P_m$, $P_1 P_2 .. \tilde P_m$, ... $\tilde P_1 \tilde P_2 ..
\tilde P_m$ are all one-dimensional projection operators and H
is of dimension $2^m$, then $P_f = f(P_1,.... P_m)$ is an
orthogonal projection on a subspace of dimension $Tr(P_f) =
wt(f)$, where $wt(f)$ is the Hamming weight of the Boolean
function $f$.
\end{thm}
\begin{proof}
By definition of $f(P_1,.... P_m)$,  we have a representation
of $P_f$ in terms of meet,  join and tilde operations in the
corresponding projection logic. By Lemma \ref{pres}, every
function of projection operators in terms of meet, join and
tilde will be present in the projection logic. Hence, $P_f$ is
an orthogonal projection operator and this proves the first
part of the theorem. Now, we will find the dimension of this
projection operator.

$f(v_1, v_2, .. , v_m)$ can be
represented as $ \sum\limits_{i = 0}^{2^m  - 1} {y_i v_1^{c_0 }
v_2^{c_1 } ....v_m^{c_{m - 1} } } $ as described in Section II. If $wt(f) = M$, then $M$
terms of $y_i$ are $1$ and the remaining terms are $0$. Also,
in this case, $P_f = f(P_1, P_2, .. , P_m) = \mathop \vee
\limits_{i = 0}^{2^m - 1} {y_i P_1^{c_0 } P_2^{c_1 }
....P_m^{c_{m - 1} } }$ (where $P_j^1 = P_j$ and $P_j^0 = \tilde P_j$). Hence, the image of $P_f$ is the
minimum subspace containing all $y_i P_1^{c_0 } P_2^{c_1 }
....P_m^{c_{m - 1}}$. We know by the statement of the theorem
that the dimension of $ P_1^{c_0 } P_2^{c_1 } ....P_m^{c_{m -
1}}$ is $1$ for all $c_0, c_1, ..., c_{m-1} \in \{0,1\}$, and all these subspaces are
orthogonal. Also, the minimum subspace containing all these
operators is the whole Hilbert space. So, the dimension of
$P_f$ will be the sum of dimensions of $y_i
P_1^{c_0 } P_2^{c_1 } ....P_m^{c_{m - 1}}$ for all $i$ (which is
$1$ when $y_i$ = $1$, and $0$ otherwise). Hence, the dimension of
$P_f$ is $M$.
\end{proof}

\vspace{0.1in} Theorem \ref{thm:bool} is a generalization of the
Theorem $1$ of \cite{Alexei2005} because we consider {\it any}
pairwise commutative projection operators, while in
\cite{Alexei2005}, a special case of commutative projection
operators using Heisenberg-Weyl group was used. This special case
is described in Section IV. Hence, to prove
Theorem \ref{thm:bool}, we use abstract properties of projection logic \cite{Cohen89}
rather than the properties of a particular commutative subgroup.
\vspace{.1in}

\begin{exa}
The Boolean function $f(v) = v_1 \bar{v}_2 + v_2 \bar{v}_3$
corresponds to the operator $P_f =f(P_1,P_2,P_3)= (P_1 \wedge
\tilde{P}_2) \oplus (P_2 \wedge \tilde{P}_3)$. If $P_1 , P_2 , P_3$
are pairwise commutative, then $P_f = P_1 + P_2 - P_1 P_2 - P_2 P_3$.
\end{exa}

\section{The Construction of Commutative Projection Operators from the Heisenberg-Weyl Group}
Let $X$, $Y$, and $Z$ be the Pauli matrices, given
by
$$
X=\left[ \begin{array}{rr}
0& 1\\
1& 0
\end{array}\right], Z=\left[\begin{array}{rr}
1&0\\
0&-1
\end{array}\right],
Y=\left[\begin{array}{rr}
0& i\\
-i& 0
\end{array}\right],
$$
 and consider linear operators $E$ of the form
$ E=e_1\otimes \ldots \otimes e_m,$ where $e_j\in\{
I_2,X,Y,Z\}. $ We form the {\it Heisenberg-Weyl
group} (sometimes in the literature this group is referred to as an extraspecial $2$-group or
as the Pauli group) $E_m$ of order $4^{m+1}$, which is realized as the group
of linear operators $\alpha E,\alpha =\pm 1,\pm i$. (For a detailed
description of the Heisenberg-Weyl group and its use to construct quantum
codes see \cite{Calderbank98}, \cite{Calderbank97}.)

Next we define the symplectic product of two vectors and the
symplectic weight of a vector.\vspace{0.1in}
\begin{defi}
The {\it symplectic inner product} of vectors $(a,b),(a',b')\in
\mathbb{F}_q^{2m}$ is given by
\begin{equation}\label{eq:symprod}
(a,b)\odot(a',b')=a\cdot b' \oplus  a'\cdot b.
\end{equation}
\end{defi}\vspace{0.1in}
\begin{defi}
The {\it symplectic weight} of a vector $(a,b)$ is the number of indices $i$
at which either $a_i$ or $b_i$ is nonzero.
\end{defi}\vspace{0.1in}

The center of the group  $E_m$ is $\{ \pm I_{2^m}, \pm i I_{2^m} \}$ and the
quotient group $\overline{E}_m$ is isomorphic to the binary vector
space $\mathbb{F}_2^{2m}$. We associate with binary vectors
$(a,b)\in \mathbb{F}_2^{2m}$ operators $E_{(a,b)}$ defined by
\begin{equation}\label{eq:Ev}
E_{(a,b)}=e_1\otimes \ldots \otimes e_m,
\end{equation}
$$
\mbox{where }e_i=\left\{
\begin{array}{ll}
I_2, & a_i=0, b_i=0,\\
X, & a_i=1, b_i=0,\\
Z, & a_i=0, b_i=1,\\
Y, & a_i=1, b_i=1.
\end{array}\right.
$$

\begin{lem}
\[
E_{(a,b)} E_{(a',b')} = (-1)^{b\cdot a'}i^{a\cdot b'+a'\cdot b}
E_{(a\oplus a',b\oplus b')}.\]
\end{lem}\vspace{0.1in}
\begin{lem}
\[
E_{(a,b)} E_{(a',b')} = (-1)^{(a,b)\odot(a',b')} E_{(a',b')}E_{(a,b)}.\]
\end{lem}\vspace{0.1in}
Thus  $E_{(a,b)}$ and $E_{(a',b')}$ commute iff $(a,b)$
and $(a',b')$ are orthogonal with respect to the symplectic inner
product (\ref{eq:symprod}).

We will now describe how to construct  commutative projection
operators. Take $m$ linearly independent vectors $y_1 , y_2 , ... ,
y_m $ of length $2m$ bits with the property that the  symplectic
product between any pair is equal to zero. If we take $ P_i  =
\frac{1} {2}(I + E_{y_i } ) $, then $P_1$, ... $P_m$ satisfy all the
properties  of Theorem \ref{thm:bool} and hence, $f(P_1 , ... P_m)$ is an
orthogonal projection operator \cite{Alexei2005}.\vspace{0.1in}
\begin{exa}
Take $f(v) = f(v_3, v_2, v_1) = v_1 + v_1 v_2 + v_3$. Take $y_1 ,
y_2$ and $y_3 $ as $(1 , 0 , 0 , 0 , 1 , 0)$ , $(0 , 1 , 1 , 1 , 1 ,
0) $ and $(0 , 0 , 1 , 0 , 1 , 1)$ respectively which are linearly independent with all pairwise
symplectic products equal to zero. Then $P_f = P_{1} \oplus P_{1} P_{2} \oplus
P_{3} = P_{1} + P_{3} - 2 P_{1}P_{3} - P_{1}P_{2} + 2
P_{1}P_{2}P_{3}$ where $P_{i} = \frac{1}{2} (I + E_{y_i})$, that is \vspace{0.1in}\[ P_f  = \frac{1} {4}\left(
{\begin{array}{*{20}c}
   2 & i & -1 & 0 & 0 & -i & 1 & 0  \\
   -i & 2 & 0 & 1 & i & 0 & 0 & -1  \\
   -1 & 0 & 2 & -i & -1 & 0 & 0 & -i \\
   0 & 1 & i & 2 & 0 & 1 & i & 0 \\
   0 & -i & -1 & 0 & 2 & i & 1 & 0 \\
   i & 0 & 0 & 1 & -i & 2 & 0 & -1 \\
   1 & 0 & 0 & -i & 1 & 0 & 2 & -i \\
   0 & -1 & i & 0 & 0 & -1 & i & 2 \\
 \end{array} } \right)
\]
\end{exa}\vspace{0.1in}

\section{Fundamentals of Quantum Error Correction}
A (($k,M$)) quantum error correcting code is an $M$-dimensional
subspace of $\mathbb{C}^{2^k}$. The parameter $k$ is the code-length
and the parameter $M$ is the dimension or the size of the code. Let
$Q$ be the quantum code, and $P$ be the corresponding orthogonal
projection operator on $Q$. (For a detailed
description, see \cite{alexquan}.) \vspace{0.1in}
\begin{defi} An
error operator $E$ is called {\it detectable} iff $PEP = c_E P$ , where
$c_E$ is a constant that depends only on $E$.
\end{defi}\vspace{0.1in}
Following \cite{EM}, we restrict attention to the errors in the
Heisenberg-Weyl group. Next, we define the minimum distance of the
code.\vspace{0.1in}

\begin{defi}
The {\it minimum distance} of $Q$ is the maximum integer $d$ such that any
error $E$, with symplectic weight at most $d-1$, is detectable.
\end{defi}\vspace{0.1in}
The parameters of the quantum error correcting code are written
(($k,M,d$)) where the third parameter $d$ is the minimum
distance of $Q$. We say that a (($k,M,d$)) quantum error correcting
code exists if there exists a (($k,M$)) quantum error correcting
code with minimum distance $\ge$ $d$. We assume $d\ge 2$ throughout the paper. We also focus
on non-degenerate (($k,M,d$)) codes, for which $PEP = 0 $ for all
errors $E$ of symplectic weight $\le$ $d-1$, which is a
sufficient condition for existence of the quantum code.

For any quantum code $Q$, we define the {\it stabilizer} $H_Q$ as
\[
H_Q  = \{ E \in E_k :E|x >  = |x > \text{ for all } |x >  \in Q\}
\]
where $E_k$ is the Heisenberg-Weyl group defined in Section IV. Then $H_Q$ is an abelian group and is isomorphic to GF$(2)^m$, for some $m$. A quantum code is called {\it additive} or a {\it stabilizer } code if it is defined by its stabilizer $H_Q$, i.e.
$$
Q = \{ |x >  \in {\Bbb C}^{2^k } :E|x >  = |x > \text{ for all } E \in H_Q \}
$$

A quantum code is non-additive if it is not equivalent to an additive code \cite{Rains99}.

\section{Quantum Error Correcting codes with minimum distance $d$}
We use $*$ to denote the standard binary inner product.
\vspace{.1in}
\begin{thm}\label{mainth}
A Boolean function $f$ with the following properties determines a $((k,M,d))$-QECC
\begin{enumerate}
\item $f$ is a function of $k$ variables and has weight $M$.
\item There are $2k$ binary $k$-tuples $x_1, x_2, ... , x_{2k}$ such that $Cset_f$ contains the set $
\{ [x_1 ,x_2 ,....x_{2k}]*w^T | $ $w$ is a 2k bit vector of
symplectic weight  $\le d-1 \}$. The rows of the matrix $A_f$ = $
\left[ {x_{1_{} } x_{2_{} } ......_{} x_{2k} } \right]_{k
\times 2k} $  have pairwise symplectic
product zero and are linearly independent.
\end{enumerate}
\vspace{0.1in} The projection operator corresponding to the
QECC is obtained as follows:
\renewcommand{\labelenumi}{(\roman{enumi})}
\begin{enumerate}
\item Construct the matrix $A_f$ as above.
\item  Define $k$ projection operators each of the form $
\frac{1} {2}(I + E_y )$ where $y$ is a row of the matrix $A_f$,
with $P_k$ corresponding to the $1^{st}$ row,  $P_{k-1}$
corresponding to the $2^{nd}$ row and so on, so that $P_1$
corresponds to the last row.
\item Transform the Boolean function $f$ into the projection operator $P_f$ using
Definition \ref{defi:bool} where  the commutative projection
operators $P_1$ .... $P_k$ are determined by the matrix $A_f$.
\end{enumerate}

\end{thm}
\begin{proof}
Consider a Boolean function $f(v)$ satisfying conditions 1) and 2). It follows easily from Section III and IV that $P_f$ constructed as above is an $M$-dimensional
projection operator. It remains to prove that the minimum distance is at least $d$, so we need to show that  $P_f \eta P_f \eta $ $=
0$ for any error $\eta$ in $E_k$ with symplectic weight at most $d-1$.

An error $\eta$ in $E_k$ transforms the projection operator $P_f$ to $P'_f = \eta P_f \eta$, and the condition $P_f \eta P_f \eta =
0$ means that $P'_f$ is orthogonal to $P_f$. Denote by $\eta_i$ the error represented by the binary $2k$-tuple with entry $1$ in position $i$ and zeros elsewhere. We emphasize that the subscripts $i$ in $x_i$, $\eta_i$ and
$A_{j,i}$ ($(j,i)^{th}$ entry
in the matrix $A_f$) are read modulo $2k$, so that  $x_{2k+1}$ is just $x_{1}$.

If $A_{1,k+1} = 0$ then $\eta_1$ commutes with $P_k$ and $\eta_1 P_k \eta_1  = P_k$, and if $A_{1,k+1} = 1$ then $\eta_1 P_k
\eta_1 = \tilde P_k$. In general, if $A_{k+1-j,k+i} = 0$ then
$\eta_i P_j \eta_i  = P_j$, and if $A_{k+1-j,k+i} = 1$ then $\eta_i
P_j \eta_i = \tilde P_j$. Let $\eta_i P_j \eta_i = Q_{i,j}$ where $Q_{i,j}$ = $P_j$ or $\tilde P_j$ and observe that $Q_{i,j} = P_j$ if and only if entry $(k+1-j)$ of $x_{k+i}$ is zero. Then $\eta_i P_f \eta_i = f(Q_{i,1}, Q_{i,2} , ... ,
Q_{i,k} )$ and the entries of $x_{k+i}$ determine $\eta_i P_f \eta_i$. In fact, this correspondence can easily be understood in terms of the fundamental correspondence between between Boolean functions
and projection operators, since the operator $\eta_i P_f \eta_i$
corresponds to the Boolean function $f(v \oplus
x_{k+i})$.

When $d=2$, we need to take care of all errors of symplectic weight $1$ by showing  $P_f \eta_i P_f \eta_i= 0$ and $P_f \eta_i \eta_{i+k} P_f \eta_i
\eta_{i+k}= 0$.
Applying the fundamental correspondence between Boolean functions and projection operators, this is equivalent to showing $f(v) f(v \oplus x_{k+i}) = 0$ and $f(v) f(v \oplus x_{k+i} \oplus x_i) = 0$ for all decimal indices $v$. This follows from the assumption that $x_{k+i}$ and $x_{k+i} \oplus x_i$ are in the complementary set $Cset_f$..

In general we need to show that $P_f \eta P_f \eta= 0$ for all errors $\eta$ of symplectic weight at most $d-1$. We write $
\eta  = \prod\limits_{i \in A} {\eta _i }
$, apply the fundamental correspondence, and find that $P_f \eta P_f \eta $ corresponds to the Boolean function $
f(v \oplus (\mathop  \oplus \limits_{i \in A} x_{i + k} ))
$. By assumption, $\mathop  \oplus \limits_{i \in A} x_{i + k}$ is in the complementary set $Cset_f$, so $f(v) f(v \oplus (\mathop  \oplus \limits_{i \in A} x_{i + k} )) = 0 $ for all $v$, and hence $P_f \eta P_f \eta= 0$.
\end{proof}
\vspace{0.1in}
Note that for $M \ge 1$ this construction only gives $((k,M,d))$ quantum error correcting codes for which the minimum distance $d$ is at most $\left\lceil {\frac{k+3}{2}} \right\rceil$. This is because any $k+1$ columns of the matrix $A_f$ are linearly dependent, which means that there is a $2k$ bit vector $w$ of symplectic weight at most $\left\lceil {\frac{k+1}{2}} \right\rceil$ such that $[x_1 ,x_2 ,....x_{2k}]*w^T = 0 $, and the zero vector is never in $Cset_f$.
\vspace{.1in}
\begin{lem}\label{addip} A $((k,M,d))$ additive QECC exists when
\begin{enumerate}
\item $M = 2^m$ for some $m$
\item There are $2k$ binary $k$-tuples $x_1, x_2, ... , x_{2k}$ such that $Cset_f$ for $f(v) = v_k v_{k-1} ... v_{m+1}$ contains the set $
\{ [x_1 ,x_2 ,....x_{2k}]*w^T | $ $w$ is a 2k bit vector of
symplectic weight  $\le d-1 \}$. The rows of the matrix $A_f$ = $
\left[ {x_{1_{} } x_{2_{} } ......_{} x_{2k} } \right]_{k
\times 2k} $  have pairwise symplectic
product zero and are linearly independent.
\end{enumerate}
\vspace{.1in}
\begin{remark}The projection operator corresponding to the QECC is $
\prod\limits_{i = m + 1}^k {\frac{1}{2}(I + E_{y_i } )}$ where $y_i$ is $k+1-i^{th}$ row of $A_f$.
The quantum code obtained in this way is that formed in the stabilizer framework using $E_{y_k}, E_{y_{k-1}}, ... , E_{y_{m+1}}$ as the stabilizers of the code.
\end{remark}
\end{lem}

\begin{proof}
By Theorem \ref{mainth} there exists a $((k,M,d))$-QECC. The construction method of Theorem \ref{mainth} gives the corresponding projection operator as $P_f =
\prod\limits_{i = m +1}^k {P_i} = \prod\limits_{i = m + 1}^k {\frac{1}{2}(I + E_{y_i } )}$. Any vector in the code subspace is given by $|x> = P_f |u>$ for some $|u> \in H$. Since $E_{y_i}$ and $E_{y_j}$ are commutative, we have $E_{y_i} |x> = |x>$ for $m< i \le k$. Hence, $E_{y_k}, E_{y_{k-1}}, ... , E_{y_{m+1}}$ are the stabilizers of the quantum code and the quantum code is additive.
\end{proof}
\begin{remark} If the boolean function can be represented as a single monomial, it gives an additive code. The converse is not true in general; see for example, \cite{Rains99}, where it is shown that every $((4,4,2))$ code is equivalent to an additive code.
\end{remark}
\vspace{.2in}

\begin{exa} For $m\ge2$,  we construct a $((2m, 4^{m-1}, 2))$ additive QECC as an example of
the above approach.  Note that Rains \cite{Rains99} has shown that
$M \le 4^{m-1}$ for any $((2m, M, 2))$ quantum code and this
example meets the upper bound. Take $f(v)$ = $v_{2m} v_{2m-1}$. It
is a function of $k=2m$ variables  with Hamming weight $4^{m-1}$
and the corresponding complementary set is $\{ (0 1 0 .. 0), (0 1
0 ... 0 1), .... (1 1 1 ... 1)\}$ (or $\{ 4^{m-1}, 4^{m-1} + 1,
.... , 4^m - 1\}$ in decimal notation). This complementary set
contains the set $\{ x_1, x_2, ... , x_{2k}, x_1 \oplus x_{k+1}, ... , x_k \oplus
x_{2k} \}$ where $x_1$ = $x_2$ = ... = $x_k$ = (0 1 0 .. 0) (or
$4^{m-1}$ ), $x_{k+1}$ = (1 0 1 .. 1), $x_{k+2}$ = (1 0 1 0 .. 0),
$x_{k+3}$ = (1 0 0 1 0 .. 0), .. , $x_{2k-1}$ = (1 0 0 .. 0 1) and
$x_{2k}$ = (1 0 0 .. 0). The matrix $A_f$ is given by
\[{\begin{array}{*{20}c}
& & & x_1 & \ldots & x_k \\
 \end{array} \begin{array}{*{20}c}
   & & &  &  \ldots & & & & x_{2k}\\
 \end{array} }
\]
\vspace{-.25in}
\[
A_f  = \left( {\begin{array}{*{20}c}
   0 &  \ldots  & 0  \\
   1 &  \ldots  & 1  \\
   0 &  \ldots  & 0  \\
    \vdots  &  \ddots  &  \vdots   \\
    0  &  \cdots  & 0  \\
   0  &  \cdots  & 0  \\
   0  &  \cdots  & 0  \\

 \end{array}| \begin{array}{*{20}c}
   1 & 1  & 1 & \ldots  & 1 & 1  & 1\\
   0 & 0  &  0 & \ldots  & 0 & 0  & 0\\
   1 & 1  &  0 & \ldots  & 0 & 0  & 0 \\
    \vdots  & \vdots  &  \vdots  &  \ddots  &  \vdots & \vdots & \vdots \\
    1 & 0  &  0 & \ldots  & 0 & 0 & 0  \\
     1 & 0  &  0 & \ldots  & 1 & 0 & 0  \\
    1 & 0  &  0 & \ldots  & 0 & 1 & 0  \\

 \end{array} } \right)
\]

\vspace{.1in}

We see that the symplectic inner product of any two rows is zero.
Hence, we have constructed a $((2m, 4^{m-1}, 2))$ QECC. Tracing
through the construction of the projection operator $P_f$ we find
that $P_f = P_{2m} P_{2m-1} $, where $ P_i = \frac{1}{2}(I +
E_{v_i} )$ and $v_i$ is the $(2m+1-i)^{th}$  row of the matrix
$A_f$. Hence, $P_{2m} = \frac{1}{2}(I + E_{00..0|11..1} )$ and
$P_{2m-1} = \frac{1}{2}(I + E_{11..1|00..0} )$.
\end{exa}\vspace{0.1in}

\begin{exa} For $m\ge3$, we construct a $((2m, 4^{m-1}, 2))$ QECC that is not additive as
 an example of the above approach. Consider the Boolean
  function $f(v) = v_{2m} v_{2m-1} v_{2m-2} + v_{2m} v_{2m-1}
  \bar{v}_{2m-2} (v_{2m-3} + \bar{v}_{2m-3} v_{2m-4} + \bar{v}_{2m-3}
  \bar{v}_{2m-4} v_{2m-5} + ... + \bar{v}_{2m-4} \bar{v}_{2m-3} ...
  \bar{v}_{2} v_{1} ) + v_{2m} \bar{v}_{2m-1} v_{2m-2} ... v_{1} $.
  It is a function of $k=2m$ variables with weight $4^{m-1}$, and the corresponding
   complementary set is
   $\{ (0 1 1 .. 1), (1 0 0 ... 0), (1 0 0 ... 1) , .... (1 1 1 ... 1)\}$
    (or $\{ 2^{2m-1}-1,$ $2^{2m-1},$ $...., 4^m - 1\}$ in decimal notation).
     This complementary set contains the set $\{ x_1, x_2, ..., x_{2k}, x_1 \oplus x_{k+1}, ... , x_k \oplus x_{2k} \}$ where $x_1$ = $x_2$ = ... = $x_k$ = (0 1 1 .. 1) (or
     $2^{2m-1} - 1$), $x_{k+1}$ = (1 0 1 .. 1), $x_{k+2}$ =
     (1 0 1 0 .. 0), $x_{k+3}$ = (1 0 0 1 0 .. 0), .. , $x_{2k-1}$ =
     (1 0 0 .. 0 1)  and $x_{2k}$ = (1 0 0 .. 0). The matrix
$A_f$ is given by
\[
 {\begin{array}{*{20}c}
   & & & x_1 & \ldots & x_k \\
 \end{array} \begin{array}{*{20}c}
    & & & & \ldots & & &  & x_{2k}\\
 \end{array} }
\]

\vspace{-.3in}

\[
A_f  = \left( {\begin{array}{*{20}c}
   0 &  \ldots  & 0  \\
   1 &  \ldots  & 1  \\
   1 &  \ldots  & 1  \\
    \vdots  &  \ddots  &  \vdots   \\
    1  &  \cdots  & 1  \\
   1  &  \cdots  & 1  \\
   1  &  \cdots  & 1  \\

 \end{array}| \begin{array}{*{20}c}
   1 & 1  & 1 & \ldots  & 1 & 1  & 1\\
   0 & 0  &  0 & \ldots  & 0 & 0  & 0\\
   1 & 1  &  0 & \ldots  & 0 & 0  & 0 \\
    \vdots  & \vdots  &  \vdots  &  \ddots  &  \vdots & \vdots & \vdots \\
    1 & 0  &  0 & \ldots  & 0 & 0 & 0  \\
     1 & 0  &  0 & \ldots  & 1 & 0 & 0  \\
    1 & 0  &  0 & \ldots  & 0 & 1 & 0  \\

 \end{array} } \right)
\]

\vspace{.1in}

     We can also see
     that the second property is satisfied, so we have constructed a
     $((2m, 4^{m-1}, 2))$ QECC  that is not additive.
\end{exa}\vspace{0.1in}
\begin{exa} The $((5, 6, 2))$-QECC constructed by Rains {\it et al.} \cite{Rains97} is also
 a special case of the above procedure. Take the Boolean function
 $f(v) = v_1 v_2 v_3 \oplus v_3 v_4 v_5 \oplus v_2 v_3 v_4 \oplus v_1 v_2 v_5 \oplus  v_1 v_4 v_5 \oplus
 v_2 v_3 v_4 v_5$. It is a function of $5$ variables with weight $6$, and the corresponding
  complementary set is $\{ 1, 3, 4, 6, 8, 11, 12, 14, 17, 19, 21, 22, 24, 26, 28, 31\}$.
  Take ($x_1$, ... , $x_{10}$ ) to be (6, 12, 24, 17, 3, 14, 31, 28, 26, 22)
  and form the matrix

\[
A_f  = \left( {\begin{array}{*{20}c}
   {0} & {0} & {1} & {1} & {0} & {0} & {1} & {1} & {1} & {1}   \\
   {0} & {1} & {1} & {0} & {0} & {1} & {1} & {1} & {1} & {0}   \\
   {1} & {1} & {0} & {0} & {0} & {1} & {1} & {1} & {0} & {1}   \\
   {1} & {0} & {0} & {0} & {1} & {1} & {1} & {0} & {1} & {1}   \\
   {0} & {0} & {0} & {1} & {1} & {0} & {1} & {0} & {0} & {0}   \\
 \end{array} } \right)\]
\vspace{.1in}

 The symplectic inner product of any two rows is zero and the
 corresponding projection operator $P_f$ coincides with  the one determined by the $((5,6,2))$-QECC
 in  \cite{Rains97}.

\end{exa}\vspace{0.1in}

\begin{lem}\label{exte}
\begin{enumerate}
     \item If there exists a $((k,M,2))$ QECC, then there exists a $((k+2,4M,2))$ QECC determined
     by $f'(v_1, v_2, ... , v_{k+2})=f(v_1, v_2, ... , v_{k})$ and
    $
    A_{f'} =(
x_1, x_2, \ldots ,x_{k-1}, x_k, x_k, x_k, x_{k+1},  x_{k+2},
\ldots ,x_{2k-1}$, $2^{k+1}+2^k+x_{2k},$ $2^k + x_{2k}, 2^{k+1} +
x_{2k})
    $
 \item If there exists a $((k,M,2))$ QECC, then there exists a $((k,M-1,2))$
    QECC determined by same $A_f$ and $f'(v)$ having support 
    a subset of
    $f(v)$.
\end{enumerate}
\end{lem}
\begin{proof}
\begin{enumerate}
\item Let $f(v_1, v_2, .., v_k)$ be the weight $M$ Boolean function corresponding to the $((k,M,2))$-QECC. The Boolean function $f'(v_1, v_2, ... , v_{k+2})=f(v_1, v_2, ... , v_{k})$ has weight $4M$, and the complementary set $Cset_{f'}$ has vectors of length $k+2$ which are of the form $\{(\{0,1\}, \{0,1\},  x):x \in Cset_f\}$. This means that  $Cset_{f'}$ has $4$ times as many elements as $Cset_{f}$. Note that if $x_1, x_2, ... , x_{2k}, x_1 \oplus x_{k+1}, ... , x_{k} \oplus x_{2k}$ are in $Cset_f$, then $(0, 0, x_1), (0, 0, x_2),$ $... (0, 0, x_{2k-1}),$ $(1, 1, x_{2k}),$ $(0, 1, x_{2k}),$ $(1, 0, x_{2k}),$ $(0, 0, x_1\oplus x_{k+1}),$ $...$, $(0, 0, x_{k-1}\oplus x_{2k-1}),$ $(1, 1, x_k \oplus x_{2k}),$ $(0, 1, x_k \oplus x_{2k}),$ $(1, 0, x_k \oplus x_{2k})$ are in $Cset_{f'}$. Let $
    A_{f'} =(
(0, 0, x_1),$ $(0, 0, x_2), $ $\ldots$ , $(0, 0, x_{k-1}),$ $(0, 0, x_k),$ $(0, 0, x_k),$ $(0, 0, x_k),$ $(0, 0, x_{k+1}),$ $(0, 0, x_{k+2}),$
$\ldots$ , $(0, 0, x_{2k-1}),$ $(1, 1, x_{2k}),$ $(0, 1, x_{2k}),$ $(1, 0, x_{2k}))
    $. All the columns and the sum of columns $i$ and $i+k$ are in $Cset_{f'}$. The symplectic product of any two rows is zero and all the rows are linearly independent, since this was true for $A_f = (
x_1 , x_2, \ldots ,x_{2k})$

\item Given this choice of $f'(v)$, we have $Cset_{f'}\supseteq Cset_f$, and this means that the same matrix $A_{f'} = A_f$ will satisfy all the earlier properties.
\end{enumerate}
\end{proof}

\vspace{0.1in}

\begin{exa} We will now use Lemma \ref{exte} to extend the Rains code to a (($2m+1, 3 \times 2^{2m-3}, 2$))-QECC
for $m > 2$.

Consider the Boolean function f(v) = $v_1 v_2 v_3 \oplus v_3 v_4 v_5 \oplus
v_2 v_3 v_4 \oplus v_1 v_2 v_5 \oplus  v_1 v_4 v_5 \oplus v_2 v_3 v_4 v_5$. It
is a function of $2m+1$
 variables with weight $3 \times 2^{2m-3}$.

 Let ($x_1$, ... $x_{2m+1}$ )  be
 (6, 12, 24, 17, 3,3,...3) and ($x_{2m+2}$, ... $x_{4m+2}$)  be
 (14, 31, 28, 26, $2^{2m+1} -10$,$2^5 +22$, $2^6 +22$, ... $2^{2m} + 22$). The matrix $A_f$  is then

\small
\[
 \left( {\begin{array}{*{20}c}
  0 & 0 & 0 & 0 & 0 & 0 &   \ldots & 0 \\
  0 & 0 & 0 & 0 & 0 & 0 &   \ldots & 0 \\

    \vdots  &  \vdots  & \vdots  & \vdots  & \vdots  & \vdots  & \ddots  &  \vdots   \\
    0 & 0 & 0 & 0 & 0 & 0 &   \ldots & 0 \\
0 & 0 & 0 & 0 & 0 & 0 &   \ldots & 0 \\
0 & 0 & 1 & 1 & 0 & 0 &   \ldots & 0 \\
0 & 1 & 1 & 0 & 0 & 0 &   \ldots & 0 \\
1 & 1 & 0 & 0 & 0 & 0 &   \ldots & 0 \\
1 & 0 & 0 & 0 & 1 & 1 &   \ldots & 1 \\
0 & 0 & 0 & 1 & 1 & 1 &   \ldots & 1 \\

 \end{array}| \begin{array}{*{20}c}
   0 & 0 & 0 & 0 & 1 & 0 &   \ldots & 1 \\
  0 & 0 & 0 & 0 & 1 & 0 &   \ldots & 0 \\

    \vdots  &  \vdots  & \vdots  & \vdots  & \vdots  & \vdots  & \ddots  &  \vdots   \\
    0 & 0 & 0 & 0 & 1 & 0 &   \ldots & 0 \\
0 & 0 & 0 & 0 & 1 & 1 &   \ldots & 0 \\
0 & 1 & 1 & 1 & 1 & 1 &   \ldots & 1 \\
1 & 1 & 1 & 1 & 0 & 0 &   \ldots & 0 \\
1 & 1 & 1 & 0 & 1 & 1 &   \ldots & 1 \\
1 & 1 & 0 & 1 & 1 & 1 &   \ldots & 1 \\
0 & 1 & 0 & 0 & 0 & 0 &   \ldots & 0 \\

 \end{array} } \right)
\]
\normalsize
\vspace{.05in}

 We see that symplectic product of any two rows is zero. Hence, we
 have constructed a (($2m+1,3\times2^{2m-3}, 2$)) non-additive QECC.

\end{exa}

\vspace{.1in}

\begin{exa}
The perfect $((5,2,3))$ additive code of R. Laflamme {\it et al.} \cite{Lafl96} can be
obtained by the above approach. Take $f(v) = v_5 v_4 v_3 v_2$. The
corresponding complementary set is $\{2,3,...31\}$. The matrix $A_f$ is given by
\[
A_f  = \left( {\begin{array}{*{20}c}
    0 & 1 & 1 & 0 & 0 & 1 & 0 & 0 & 1 & 0 \\
    0 & 0 & 1 & 1 & 0 & 0 & 1 & 0 & 0 & 1 \\
    0 & 0 & 0 & 1 & 1 & 1 & 0 & 1 & 0 & 0 \\
    1 & 0 & 0 & 0 & 1 & 0 & 1 & 0 & 1 & 0 \\
    0 & 0 & 1 & 0 & 0 & 1 & 0 & 0 & 0 & 1 \\

 \end{array} } \right),
\]
it is easy to see that all rows are linearly independent, and
that the symplectic inner product of any two rows is zero. Note that the stabilizers corresponding to the code are $ZXXZI$, $IZXXZ$, $ZIZXX$, and $XZIZX$.
\end{exa}

\section{Operator Quantum Error Correction (OQEC)}
The theory of operator quantum error correction \cite{Kribs05}
uses the framework of noiseless subsystems to improve the
performance of decoding algorithms which might help improve the
threshold for fault-tolerant quantum computation. It requires a
fixed partition of the system's Hilbert space $H = A \otimes B
\oplus C^ \bot$. Information is encoded on the A subsystem;
 the logical quantum state $\rho_A \in \Bbb B_A$ is
encoded as $\rho_A \otimes \rho_B \oplus 0^{C^ \bot}$ with an
arbitrary $\rho_B \in \Bbb B_B$ (where $\Bbb B_A$ and $\Bbb B_B$
are the sets of all endomorphisms on subsystems A and B
respectively). We say that the error $E$ is correctable on
subsystem $A$ (called the logical subsystem) when there exists a physical map $R$ that reverses
its action, up to a transformation on the $B$ subsystem (called the Gauge
subsystem). In other
words, this error correcting procedure may induce some nontrivial
action on the $B$ subsystem in the process of restoring
information encoded in the $A$ subsystem. This leads to recovery routines which
explicitly make use of the subsystem structure \cite{Bacon06}\cite{Zanardi04}.
In the case of standard
quantum error correcting codes, the dimension of $B$ is $1$. A $((k, M,
N, d))$-OQEC is defined as a OQEC in $\Bbb C^{2^k}$ with $M$ and $N$ as the dimension of the logical and gauge subsystems.

\vspace{.1in}
\begin{lem}A Boolean function $f$ with the following properties determines $((k,2^t, 2^{s-t},d))$ stabilizer OQEC
\begin{enumerate}
\item $f(v)$ is of the form $v_k v_{k-1} .. v_{s+1}$ with weight $2^s$
\item There are $2k$ binary $k$-tuples $x_1, x_2, ... , x_{2k}$ such that $Cset_f$ contains the set $
\{ [x_1 ,x_2 ,....x_{2k}]*w^T | $ $w$ is a 2k bit vector of
symplectic weight  $\le d-1 \}$. The rows of the matrix $A_f$ = $
\left[ {x_{1_{} } x_{2_{} } ......_{} x_{2k} } \right]_{k
\times 2k} $  have pairwise symplectic
product zero and are linearly independent.

\end{enumerate}
\end{lem}
\begin{proof}
By Lemma \ref{addip}, $f(v)$ satisfies the conditions for construction of an additive $((k,2^s,d))$-QECC. The first $k-s$ rows of the matrix $A_f$ are the stabilizers of the code, and using this QECC, we construct an OQEC following \cite{Poulin06}.

We denote by $X_j$ the matrix $X$ (the Pauli
matrix) acting on the $j^{th}$ qubit, and similarly for $Y_j$ and
$Z_j$. The Heisenberg-Weyl group $ E_k  =  < i, X_1, Z_1, ... , X_k, Z_k
>$. The first step in constructing a stabilizer code is to choose
a set of $2k$ operators $ \{ X_j ^\prime , Z_j ^\prime  \} _{j =
1, .. , k} $  from $E_k$ that is Clifford isomorphic to the set of
single-qubit Pauli operators $ \{ X_j, Z_j \} _{j = 1, .. , k} $
in the sense that the primed and unprimed operators obey the same
commutation relations. The operators $ \{ X_j
^\prime, Z_j ^\prime  \} _{j = 1, .. , k} $  generate $P_k$ and
behave as single-qubit Pauli operators. We can think of them as
acting on $k$ virtual qubits.

Form $Z_1 ^\prime$, ... , $Z_k ^\prime $ corresponding to the rows
of matrix $A_f$. (The image of the first row in the Heisenberg-Weyl group
gives $Z_1 ^\prime$ and so on.) Given all the $Z_j ^\prime$, we
can easily find $X_j ^\prime$ which have symplectic product of $1$
with $X_j ^\prime$ and symplectic product of $0$ with all other
$X_l ^\prime$, $ l \ne j $.

Hence, the stabilizer group is given by $ S=$  $<Z_1 ^\prime,$
$Z_2 ^\prime, ... ,$ $Z_{k-s}^\prime>$. If we want to construct a
$((k, 2^t, 2^{s-t}, d))$-OQEC, then we need to find a subsystem of
dimension $2^t$ in the above subspace $C$ of dimension $2^s$. Following \cite{Poulin06}, if we take the Gauge group (corresponding to
the Gauge subsystem defined before) $ G =$ $<S, X_{k-s + 1}
^\prime, Z_{k-s + 1} ^\prime, ... , X_{k - t} ^\prime, Z_{k - t}
^\prime> $ and the logical group  $ L =$ $<X_{k - t + 1} ^\prime,
Z_{k - t + 1} ^\prime, ... , X_k ^\prime, Z_k ^\prime> $, the
action of any $l \in L$ and $g \in G$ restricted to the code
subspace $C$ is given by
\[
\begin{array}{l}
 gP = I_A  \otimes g^B  \\
 lP = l^A  \otimes I_B  \\
 \end{array}
\]
for some $l^A$, $g^B$ in $\Bbb B_A$ and $\Bbb B_B$ respectively,
where $A$ and $B$ are the required subsystems. Since we are encoding in a subsystem of the subspace formed by $((k,2^s,d))$-QECC, the minimum distance of the OQEC thus obtained will be $\ge d$.

\end{proof}

\section{Conclusion}
We have described a fundamental correspondence between Boolean functions and projection operators in Hilbert space that provides a mathematical framework that unifies the
construction of additive and non-additive quantum codes. We have given sufficient conditions for
the existence of QECC in terms of existence of a Boolean
function satisfying certain properties and presented examples of Boolean functions satisfying these properties. We have also given a method to construct the quantum code if these properties are satisfied. Our method leads to a construction
of $((2m, 4^{m-1} , 2))$ codes, the
original $((5,6,2))$ code constructed by Rains {\it et al.}, the
extension of this code to $((2m+1,3 \times 2^{2m-3},2))$ codes, and
the perfect $((5,2,3))$ code. Finally we have shown how the new
framework can be integrated with operator quantum error
correcting codes.

\section{Acknowledgements}
The authors would like to thank the anonymous reviewers for many suggestions that improved this paper and for bringing the work of Danielson \cite{thesis} to their attention.
\bibliographystyle{IEEEbib}

\end{document}